\documentclass[prl,twocolumn,showpacs,preprintnumbers,amsmath,amssymb]{revtex4}

\usepackage{graphicx}% Include figure files
\usepackage{dcolumn}% Align table columns on decimal point
\usepackage{bm}% bold math
\begin{document}

\preprint{APS/123-QED}

\title{Accurate modeling of left-handed media using finite-difference time-domain method and
finite-size effects of a left-handed medium slab on the image
quality revisited}

\author{Yan Zhao}
\email{yan.zhao@elec.qmul.ac.uk}
\author{Pavel Belov}
\author{Yang Hao}
\affiliation{Department of Electronic Engineering, Queen Mary
University of London, Mile End Road, London, E1 4NS, United Kingdom}

\date{\today}% It is always \today, today, but any date may be explicitly specified

\begin{abstract}

The letter contains an important message regarding the numerical
modeling of left-handed media (LHM) using the finite-difference
time-domain (FDTD) method which remains at the moment one of the
main techniques used in studies of these exotic materials. It is
shown that conventional implementation of the dispersive FDTD method
leads to inaccurate description of evanescent waves in the LHM. This
defect can be corrected using the spatial averaging at the
interfaces. However, a number of results obtained using conventional
FDTD method has to be reconsidered. For instance, the accurate
simulation of sub-wavelength imaging by the finite-sized slabs of
left-handed media does not reveal the cavity effect reported in
[Phys. Rev. Lett. \textbf{92}, 107404 (2004)]. Hence the finite
transverse dimension of LHM slabs does not have significant effects
on the sub-wavelength image quality, in contrary to previous
assertions.

\end{abstract}

\pacs{78.20.Ci, 42.30.Wb, 73.20.Mf\vspace{-1mm}}% PACS, the Physics and Astronomy Classification Scheme.

\maketitle

The finite-difference time-domain method (FDTD) is known as one of
the powerful numerical techniques in electrodynamics \cite{Taflove}.
Being simple in implementation it has been proved to be very popular
among researchers. This method is assumed to be extremely accurate
since it involves direct numerical solution of Maxwell equations
which are known as the basis of classical electrodynamics. However,
the implicit belief in FDTD sometimes results in attributing certain
physical properties to some electromagnetic structures based on
simulation results. One typical example is a long row of works on
FDTD modeling of the left-handed media (LHM), materials with
negative permittivity and permeability \cite{Veselago}. Such
materials are not yet available experimentally and thus numerical
simulations still remain one of the most common ways in exploring
their properties and applications. However, as it will be shown in
this letter, the conventional implementation of FDTD for modeling of
LHM in the same manner as for usual dispersive dielectric materials
leads to incorrect simulation results. It may seem that the
conventional FDTD has been verified in the literature: the negative
refraction effect which is inherent to the boundary between the free
space and LHM was observed and the planar superlens behaviour has
been successfully demonstrated
\cite{ZiolkowskiPRE,Loschialpo,Cummer}. Actually, this only means
that the LHM is correctly modeled for the case of propagating waves.
As soon as the evanescent waves are considered the conventional
implementation of the FDTD fails. Usually, the evanescent waves
decay exponentially over the distance and thus they are concentrated
in the close vicinity of sources, that is why conventional FDTD
modeling of usual materials does not suffer from this trouble. In
the case of LHM, the evanescent waves play key roles and have to be
modeled accurately because of the perfect lens effect
\cite{Pendrylens}. A slab of LHM effectively amplifies evanescent
waves which normally decay in usual materials and allows
transmission of sub-wavelength details of sources to significant
distances. Ideally lossless LHM slabs provide unlimited
sub-wavelength resolution. However in realistic situations, the
resolution is restricted by losses and the thickness of the slab
\cite{Podolskiy}, as well as the mismatch between the LHM and its
surrounding medium \cite{SmithLimit}.

\begin{figure}[t]
\centering
\includegraphics[width=8.6cm]{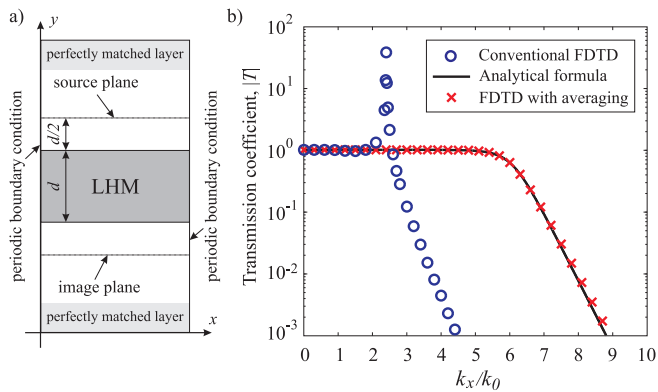}
\caption{(Color online) (a) The simulation domain for calculation of
the transmission coefficient. (b) The comparison of transmission
coefficients as functions of transverse wave vector $k_x$ for the
LHM slab with $\epsilon=\mu=-1-0.001j$ and thickness $d=\lambda/5$,
calculated using conventional FDTD method (circles)  with the
analytical result (solid line). The result provided by the scheme
with averaging at the boundaries is given by crosses.\vspace{-4mm}}
\label{fig_transmission}
\end{figure}

In order to illustrate what we mean by incorrect description of
evanescent waves provided by conventional implementation of FDTD, we
have simulated propagation of plane electromagnetic waves (with TM
polarization and various transverse wave vectors) through an
infinite slab of LHM in the free space. We choose this problem since
an analytical expression for transmission coefficient through the
LHM slab is available \cite{Podolskiy}, that allows us to check
validity of our numerical results. The LHM slabs with relative
permittivity and permeability $\varepsilon=\mu=-1-0.001j$ and
thickness $d=\lambda/5$, where $\lambda$ is wavelength in the free
space, are tested. The frequency dispersion of the LHM has been
modeled by the Drude model with the plasma frequency
$\omega_p=\sqrt{2}\omega$ and the collision frequency
$\gamma=0.0005\omega$, where $\omega$ is the operating frequency. It
is implemented in FDTD using the auxiliary differential equation
(ADE) method \cite{Taflove}. The two-dimensional (2-D) simulation
domain is bounded by perfectly matched layers (PMLs) and periodical
boundary conditions (PBCs) as illustrated in
Fig.~\ref{fig_transmission}(a).
A soft current sheet source (which allows scattered waves to pass
through) with phase delay corresponding to different transverse wave
vectors $k_x$ placed at the distance $d/2$ from the slab is used as
excitation. The PBCs are also specified by the particular wave
vector $k_x$. The source is slowly and smoothly switched to its
maximum value in order to avoid exciting other frequency components
\cite{ZiolkowskiPRE}. The computation continues until the steady
state is reached. The simulation has been done for both propagating
($k_x<k$, where $k$ is wave vector in the free space) and evanescent
($k_x>k$) waves. The Berenger's original PMLs \cite{Berenger} are
used for absorbing propagating waves, and the modified PMLs
\cite{FangGPML} are applied when the transmission coefficient for
evanescent waves is calculated. The PMLs are placed at $\lambda/2$
distance away from the slab. We use Yee's square grid with periods
$\Delta x=\Delta y=\lambda/100$ and the time step $\Delta t=\Delta
x/\sqrt{2}c$, where $c$ is the speed of light in the free space,
chosen according to the Courant stability condition \cite{Taflove}.
Since infinite structures can be truncated with any period, in order
to save computation time, we use only four FDTD cells along
$x$-direction.

The calculated transmission coefficient from the source plane to the
image plane (located at the distance $d/2$ from the other side of
the slab) as function of $k_x$ is presented in
Fig.~\ref{fig_transmission}(b) by circles. The reference curve
calculated using the analytical expression \cite{Podolskiy} is also
shown for comparison. It is clearly visible that the numerical
results from the conventional FDTD are correct only for $k_x<2k_0$.
This range of $k_x$ covers all propagating waves ($k_x<k_0$) and a
small part of weakly decaying evanescent waves ($k_0<k_x<2k_0$). For
the evanescent waves with $k_x>2k_0$ the numerical results
dramatically differ from the analytical results and the former shows
resonant behavior with a strong peak at $k_x=2.4k_0$. This effect
can be explained as resonant excitation of a `numerical surface
plasmon' at the back interface of LHM slab. Similar phenomena can be
observed in the case of metallic slabs \cite{Pendrylens} or for
unmatched LHM \cite{SmithLimit}, but in this particular case it is
purely numerical artefact. The incorrect behavior of numerical
solutions remains similar if the FDTD grid period is reduced to
$\lambda/200$ and $\lambda/400$, but the resonance shifts to
$k_x=2.8k_0$ and $k_x=3.2k_0$, respectively.

The presence of `numerical surface plasmons' provides evidence that
the boundaries between the LHM and the free space have not been
modeled accurately. If at the boundaries the mean value of
permittivity of LHM and the free space is used for updating the
tangential component of electric field (which is equivalent to the
spatial averaging suggested in \cite{LimitFDTD}) then the spurious
`numerical surface plasmons' disappear and the modeling happens to
be extremely accurate. The transmission coefficient calculated using
the proposed spatial averaging at the boundaries are presented in
Fig.~\ref{fig_transmission}(b) by crosses. It is clear that the
result repeats the estimated analytical values with very good
accuracy for the whole spatial spectrum of waves. The calculation
has been performed for $\Delta x=\Delta y=\lambda/100$, and it
remains accurate even for larger grid periods. The above simple test
allows us to conclude, that the conventional FDTD method fails to
describe the propagation of high-order evanescent waves in the LHM
if no corrections at the boundaries of the LHM are made. As a
result, a number of previously obtained results using the
conventional FDTD have to be reconsidered. This especially concerns
the modeling of sub-wavelength imaging by LHM slabs
\cite{Pendrylens} which involves operation with evanescent waves.
Note, that the simulations where only propagating waves are involved
(e.g. demonstration of negative refraction for an obliquely incident
plane wave) are not affected by this problem.

The numerical transmission coefficient for LHM slabs reported in
\cite{Rao} is a typical example of using the FDTD method without
averaging. The study can actually be treated as an investigation of
the `numerical surface plasmons', their sensitivity to losses and
efficiency of their excitation for various thicknesses of the slab.
Unfortunately, these results have no relations with the properties
of actual LHM slabs. The performance of the LHM slab as a
sub-wavelength imaging device indeed depends on the losses and the
thickness of the slab as it is shown in \cite{Podolskiy}, but it is
completely different dependence as compared to results reported in
\cite{Rao}.

\begin{figure}[t]
\centering
\includegraphics[width=8.6cm]{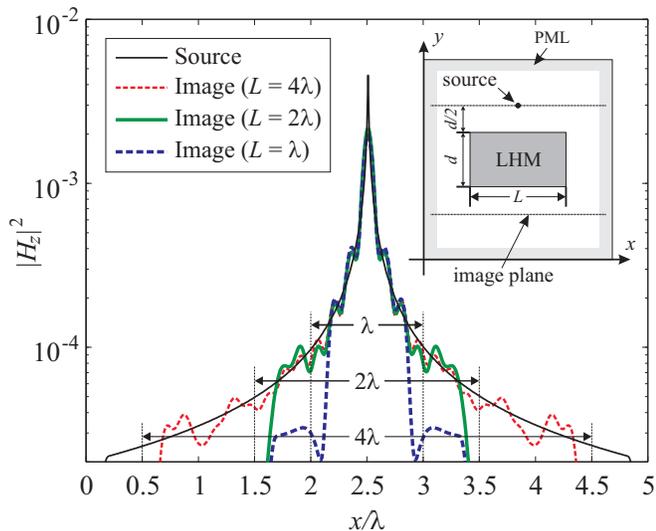}
\caption{(Color online) The magnetic field intensities at the image
planes of the planar LHM lenses ($\varepsilon=\mu=-1-0.001j$,
$d=\lambda/5$) with various transverse dimensions: $L=\lambda$,
$2\lambda$ and $4\lambda$.\vspace{-4mm}} \label{fig_imaging1}
\end{figure}

One of the recent most puzzling results related to the quality of
imaging provided by LHM slabs is reported in \cite{SailingFinite}.
It is claimed that the operation of the finite-sized structures is
significantly affected by their transverse dimensions. Having the
FDTD code with spatial averaging at the interfaces which has been
proven to be accurate, we decide to check this statement. The
finite-sized slabs of LHM excited by magnetic current sources are
modeled for three different transverse dimensions: $L=\lambda$,
$2\lambda$ and $4\lambda$, as illustrated by the sketch in
Fig.~\ref{fig_imaging1}. The parameters of the LHM slab
($\varepsilon=\mu=-1-0.001j$, $d=\lambda/5$) and distance between
the source and the front interface equal to $d/2$ are kept the same
for all simulations. The simulation domain is truncated by PMLs
located at $\lambda/2$ distance away from the LHM slab and both
source and image planes. The same grid periods $\Delta x=\Delta
y=\lambda/100$ and the time step $\Delta t=\Delta x/\sqrt{2}c$ are
used as for previous plane-wave simulations. The source is switched
slowly and smoothly in order to avoid contributions from undesired
frequency components \cite{ZiolkowskiPRE} and the simulations last
until the steady state is reached. The intensity distributions in
the image planes for all three cases of different transverse
dimensions are plotted in Fig.~\ref{fig_imaging1}.
It is clear that the image quality is practically unaffected by the
transverse size of the slab. The distributions repeat the source
distribution, which is plotted in the same figure, with good
sub-wavelength resolution. We do not observe any distortion of
images caused by the finiteness of the transmission device, in
contrast to the conclusions made in \cite{SailingFinite}. The slight
disagreement between the image and the source is due to the finite
resolution of lenses caused by the losses in LHM. We suppose that
the resonant effects and image distortions related to transverse
dimensions reported in \cite{SailingFinite} can be interpreted as
excitation of `numerical surface plasmons' at the interfaces of the
slab and can be observed only in inaccurate FDTD simulations without
spatial averaging at the boundaries. Thus, these effects are purely
numerical and have no relation to the properties of actual LHM
slabs. In reality, the imaging performance of finite-sized LHM slabs
is unaffected by their transverse dimensions. This statement also
has been confirmed by full-wave electromagnetic simulation using
Ansoft HFSS$^{\rm TM}$ package.

\begin{figure}[t] \centering
\includegraphics[width=8.6cm]{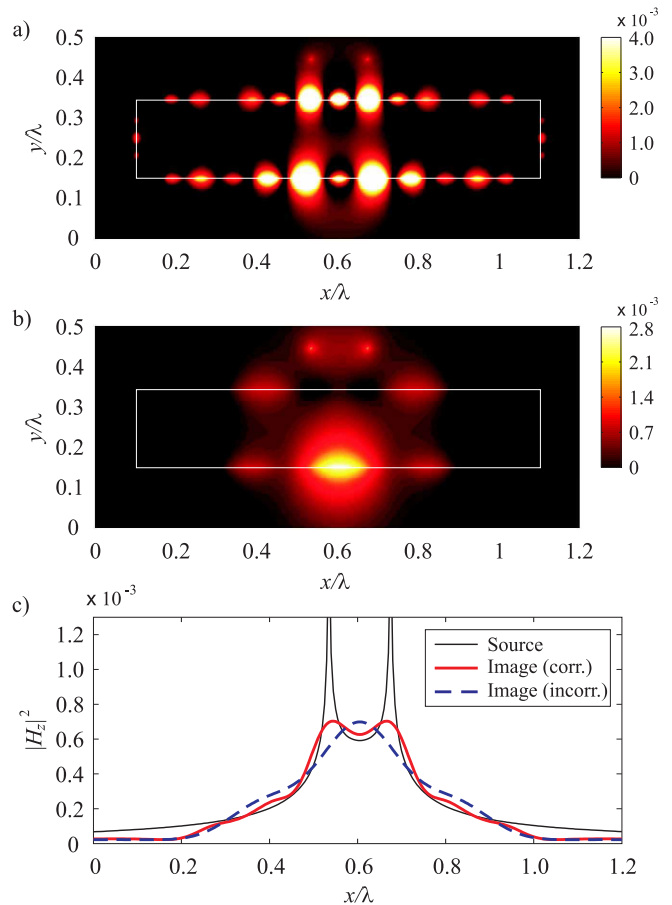}
\caption{(Color online) The magnetic field intensity distributions
around the LHM slab with transverse size $L=\lambda$ excited by two
magnetic line sources placed at $\lambda/8$ distance between each
other obtained using FDTD method (a) with and (b) without spatial
averaging. (c) The image and source plane cuts.\vspace{-4mm}}
\label{fig_imaging2}
\end{figure}

In order to illustrate this statement we have performed the
simulation for a LHM slab with transverse size $L=\lambda$ excited
by two magnetic line sources placed at $\lambda/8$ distance between
each other using the FDTD method with spatial averaging at the
boundaries. The rest of parameters is the same as before. The
distance between the sources is larger than the resolution of the
lens which should be better than $\lambda/12$ based on the fact that
the transfer function plotted in Fig.~\ref{fig_transmission} is
close to unity for $k_x/k<6$. This allows us to expect two
well-resolved maxima in the image plane. The distribution of
magnetic field intensity in a sub-domain near the LHM slab (the
actual FDTD domain is larger) in the steady state is presented in
Fig.~\ref{fig_imaging2}(a).
The distribution in the image plane is shown in
Fig.~\ref{fig_imaging2}(c) together with the source. Two maxima at
the distance of $\lambda/8$ are clearly visible in the image plane.
This confirms sub-wavelength imaging capability of the LHM lens with
only one wavelength width. Thus, we do not observe any limitations
on the functionality of the LHM slabs as sub-wavelength imaging
devices due to their finite transverse dimensions. However, if the
same system is modeled by FDTD method without corrections at the
boundaries then completely different distribution of the field
around the slab is observed, see Fig.~\ref{fig_imaging2}(b). The
distribution of the field along the slab interface is smooth which
once again confirms that high-order evanescent waves are not
correctly modeled. As a result, the sub-wavelength details of the
source are not resolved in the image plane. Instead of expected two
closely located maxima the intensity distribution in the image plane
has only one wide maximum, see Fig.~\ref{fig_imaging2}(c).

We suppose that the presented comparison between FDTD models with
and without spatial averaging at the boundaries clearly demonstrates
the limitation of the conventional FDTD method for modeling of LHM
lenses. The significant discrepancies appear only in the cases when
evanescent waves are involved. However, we encourage to use the
model with corrected updating equations at the boundaries (with
spatial averaging of permittivity as we propose or with averaging of
the current as proposed in \cite{LimitFDTD} which are equivalent) in
all cases in order to avoid numerical artefacts. The spatial
averaging is known as a second-order correction in the case of
boundaries between dielectrics with positive permittivity
\cite{Hwang}, but in the case of evanescent waves in LHM it
transforms into essential and mandatory correction.

In addition to the corrections at the boundaries we would like to
stress a few other aspects which are important to accurate FDTD
modeling of LHM. The numerical dispersion is usually assumed to have
very little influence on the quality of FDTD simulation. However,
the properties of LHM are known to be extremely sensitive to minor
changes of material parameters. A typical example is the degradation
of resolution due to small variations of material parameters from
$\varepsilon=\mu=-1$ \cite{Podolskiy,SmithLimit}. This means that
even a tiny amount of numerical dispersion may cause crucial
discrepancy between simulated and theoretical results. Fortunately,
the numerical dispersion can be easily analyzed analytically and an
estimation for the difference between effective numerical material
parameters and the `real' ones are known \cite{LimitFDTD}. This
allows to adjust parameters of the FDTD simulation to ensure that
the effective numerical material parameters correspond to required
values. We suggest using this adjustment in the cases of large time
steps and small losses.

It is well known that the switching time considerably influences the
oscillation of images created by LHM lenses. The switching time
equal to thirty periods was used in \cite{Cummer,Rao}. However,
perhaps this is the reason why no stable images could be obtained in
\cite{ZiolkowskiPRE,Rao}. Recently, it was reported in
\cite{HuangSwitching} that a switching time equal to one hundred
periods is required to obtain stable images. We have used such a
switching time in all our simulations and we recommend to pay
attention to the issue of switching time in all FDTD simulations of
LHMs. The high-order evanescent waves travel very slowly in the LHM
slabs and the procedure of the amplification of evanescent waves
requires very long time to reach the steady-state. That is why, in
addition to smooth and slow switching of the source we recommend to
add losses into the simulations in order to limit the spatial
spectrum of evanescent waves which are involved in operation.
Otherwise, in the lossless case the transient processes may last
extremely long.

The diffraction on the wedges and corners also may provide certain
problems because of the singularity effects reported in
\cite{LHMCorner}. However, in the case of LHM with
$\varepsilon=\mu\approx-1$ the singular behavior disappears and the
wedges operate mainly as retroreflectors \cite{CornerMonzon}. That
is why, in the simulations presented in this letter we have not
observed any singularities at the wedges.

In conclusion, we have demonstrated in the letter that the
conventional FDTD method for modeling of LHM leads to inaccurate
description of high-order evanescent waves and does not simulate
sub-wavelength imaging correctly. The simulations suffer from the
artefact of `numerical surface plasmons' which appear at the
interfaces of LHM. In order to solve this problem and ensure
accurate FDTD modeling, a spatial averaging scheme at the boundaries
has been proposed in the same manner as in \cite{LimitFDTD}. This
technique has been tested on the analytically solvable example of a
plane wave propagation through an infinite LHM slab and extremely
good accuracy of simulation has been revealed for all angles of
incidence including high-order evanescent waves. The finite-sized
slabs of LHM excited by a single line source have been modeled using
the proposed technique for various transverse dimensions of the slab
and the sub-wavelength imaging capability of the structures has been
confirmed. No distortions of the image due to the finite transverse
dimensions of the structure have been revealed, in contrast to the
results reported in \cite{SailingFinite}. We suspect that the
resonant effects due to finite transverse dimensions of the LHM slab
reported in \cite{SailingFinite} are caused by excitation of the
`numerical surface plasmons' and are completely numerical. These
effects are absent in real structures and there are no restrictions
on functionality of LHM sub-wavelength lenses due to their
transverse dimensions. Our FDTD simulations with spatial averaging
at the boundaries demonstrate the imaging of two line sources
located at sub-wavelength distance between each other by a LHM slab
with only one-wavelength width. Keeping in mind that simulations
using the FDTD method remain one of the most popular approaches in
the studies of LHM, the general recommendations on how to accurately
model LHM using the FDTD method are also provided.

\bibliography{LHM-PRL}% Produces the bibliography via BibTeX.

\end{document}